\documentclass{moriond}

\usepackage{cancel}
\def\Journal#1#2#3#4{{#1} {\bf #2}, #3 (#4)}

\def\NPB{{\em Nucl. Phys.} B}
\def\PLB{{\em Phys. Lett.}  B}

\def\PRD{{\em Phys. Rev.} D}

\def\be{\begin{equation}}
\def\ee{\end{equation}}
\def\bea{\begin{eqnarray}}
\def\eea{\end{eqnarray}}

\newcommand{\Photo}{}

\begin{document}
\vspace*{4cm}
\title{SEARCH FOR A SM-LIKE HIGGS BOSON IN THE H$\rightarrow$ZZ$\rightarrow\ell\ell q\bar{q}$ DECAY CHANNEL IN CMS}

\author{ E. NAVARRO DE MARTINO }

\address{Dpto. de Investigaci\'{o}n B\'{a}sica, Av. Complutense, 40,\\
CIEMAT, Madrid, Spain}

\maketitle\abstracts{
A search for a high mass standard-model-like Higgs boson decaying into two Z bosons with subsequent decay into two leptons and two quarks performed at CMS is presented. The analysis is based on 19.7 $\mathrm{fb}^{-1}$ of proton-proton collisions produced in LHC at $\sqrt{s}=$~8~TeV. Different categories are exploited in order to isolate hypothetical Higgs boson-like signals in the mass range up to 1~TeV. The data are interpreted in terms of a standard-model-like Higgs boson as well as an electroweak singlet, visible through the interference with the 125~GeV Higgs boson. No evidence of a signal is found and upper limits are set on the production cross section and other model parameters. 
}

\section{Introduction}

The recent discovery of a scalar particle in the CMS and ATLAS experiments, h(125), with properties compatible with those predicted for the SM Higgs boson, opens the possibility of the existence of additional particles belonging to an enlarged scalar sector. In this context, CMS has exploited the decay channels of a heavy Higgs boson, H, decaying into a pair of gauge bosons, H$\rightarrow$ZZ and H$\rightarrow$WW. This work reports the search for such a heavy Higgs boson in the decay mode H$\rightarrow$ZZ$\rightarrow\ell\ell q\bar{q}$~\cite{ref:2l2q}. The analysis uses data from proton-proton collisions at $\sqrt{s}=$~8~TeV, collected by CMS during 2012, corresponding to an integrated luminosity of 19.7~fb$^{-1}$.

The process H$\rightarrow$ZZ$\rightarrow\ell\ell q\bar{q}$ shows a clear signature in CMS, consisting of two high-momentum leptons and two high-momentum hadronic jets. The analysis suppresses the overwhelming background contribution, which has a cross section larger than $10^3$ times that of the H boson. The dominant background is the production of a Z boson in association with one or more jets, Z+jets. Other relevant backgrounds stem from top quark pairs, decaying into two leptons, two b-quarks and 2 neutrinos, and the direct production of gauge boson pairs (ZZ, WW and WZ). 

The mass of the two reconstructed leptons and the two hadronic jets, $m_{\ell\ell q\bar{q}}$, is used as the main discriminant between signal and background. This observable is completely reconstructed since all products are detected in CMS, allowing for a powerful discrimination. For masses above 200~GeV, the $m_{\ell\ell q\bar{q}}$ distribution has an exponentially falling shape, while a high mass Higgs boson signal would present the form of a resonance, centered in the H mass, $m_H$. The two main production mechanisms are considered in this work, gluon fusion (ggH) and vector boson fusion (VBF).

\section{Data analysis}\label{sec:analysis}

To reconstruct a heavy Higgs boson decaying into H$\rightarrow$ZZ$\rightarrow\ell\ell q\bar{q}$, events with two opposite-sign same-flavor leptons and two hadronic jets are selected. They are recorded in CMS using triggers requiring two high-momentum leptons, with transverse momentum $p_{T}>$17(8)~GeV for the first(second) lepton.

Isolated leptons (electrons or muons), $\ell$, are used to build leptonic Z boson candidates, Z$\rightarrow\ell^{+}\ell^{-}$. To ensure high efficiency and good momentum determination the leading (subleading) lepton is required to have $p_{T}>$40(20)~GeV. Muons inside the pseudorapidity range $|\eta|<2.4$ are considered for analysis, while electrons are selected in the $|\eta|<2.5$ range, excluding the transition region between the barrel and the endcap, $1.44<|\eta|<1.57$. Each Z$\rightarrow\ell^{+}\ell^{-}$ candidate is restricted to the dilepton mass range 76~GeV$<m_{\ell\ell}<$106~GeV in order to reduce backgrounds not containing a real leptonic Z, such as the top quark pair production. 

Hadronic jets, j, are reconstructed by clustering particles with the anti-$k_T$ algorithm~\cite{ref:ak5}, with radius parameter R=0.5 (AK5 jets). To avoid double counting, AK5 jets overlapping with a reconstructed lepton are removed from the event. Each pair of jets is considered as a hadronic Z boson. The energy of the jet is corrected to account for the non-linear response of the calorimeter and detector noise. To ensure good momentum determination, AK5 jets are selected for analysis if they have $p_{T}>$30~GeV, after energy corrections. For high resonance masses, above 600~GeV, both quarks emerging from the collision are highly boosted. Therefore, the two jets originating from the quarks overlap. In this case, the requirement of having two distinct jets in the event leads to a drastically lost of efficiency. To recover the events with boosted jets, two leptons and a single merged jet, J, are required in the events. The merged jet is reconstructed using the Cambridge-Aachen algorithm~\cite{ref:ak5} with radius parameter 0.8 (CA8 jets). Events with $p_T(\ell^+\ell^-)>200$~GeV and a CA8 jet with $p_T(J)>$100~GeV are considered as boosted candidates. To improve the mass resolution a jet pruning algorithm is performed on the CA8 jets~\cite{ref:boost}. Additionally, the boosted candidates are required to have n-subjettiness ratio~\cite{ref:boost} of $\tau_{21}<0.5$, in order to reject background from the hadronization of single quarks and gluons.

The dijet mass is restricted to the 71~GeV$<m_{jj}<$111~GeV region, largely reducing the Z+jets background. Additionally, a sideband region is defined for control purposes and background determination, requiring events with $m_{jj}$ in the $[60,71]\cup[111,130]$~GeV region.

Since the Higgs boson is a scalar particle that carries no spin, five angles fully describe the decay kinematics~\cite{ref:gao}. A single angular discriminant is built from the probability ratio of the signal and background angular distributions. This discriminant is weakly correlated with the ZZ, $\ell\ell$ and $q\bar{q}$ masses, and provides a good discrimination power between background and signal.

Jets from the hadronic Z decay have higher b-quark content than jets coming from the Z+jets background. Thus, the parton flavor identification of the jets (b-tagging) is exploited to enhance the sensitivity to the signal. To identify jets coming from the b-quark hadronization, the Jet Probability algorithm~\cite{ref:btag} is used, either on both AK5 jets or on the subjets identified from the CA8 merged jet.

The $\mathrm{t\bar{t}}$ background is especially relevant in events with two b-tagged jets, characterised by a significant amount of missing transverse energy, $\cancel{E}_{T}$. Signal events are, however, typically well-balanced. The $\cancel{E}_{T}$ significance~\cite{ref:2l2q}, $\lambda$, is defined as the likelihood that the observed $\cancel{E}_{T}$ in an event comes from true $\cancel{E}_{T}$ and not from detector-related limitations. The loose requirement $\lambda<10$ largely reduces the $\mathrm{t\bar{t}}$ contamination.

The hadronic and leptonic reconstructed Z bosons are used to reconstruct Higgs boson candidates. To obtain the best sensitivity to the signal, the H candidates are classified into 14 exclusive categories, according to the following criteria. First, the event is separated by its production mechanism. If the event contains two forward AK5 jets ($|\eta|<4.7$) with a pseudorapidity difference $\Delta\eta>3.5$ and high invariant mass, $m_{jj}>500$~GeV, it is identified as a VBF-like event. Otherwise, it is considered as a ggH-like event. The event sample is also split into the dielectron and dimuon categories. Events are further classified by the hadronic Z decay: if the event contains a suitable CA8 jet passing the $p_{T}>100$~GeV and $\tau_{21}<0.5$ criteria, it is categorized as a merged jet event. The remaining events constitute the dijet category. The ggH-like events are additionally split into three categories, according to the number of b-tagged (sub)jets: the 0, 1, 2 btag categories.

\section{Background and signal modeling}

The main background of the analysis is Z+jets. Its shape is extracted from large simulated samples, generated using MadGraph. The sideband control region mentioned in Sec.~\ref{sec:analysis} is used to constrain its relative normalization. The control sample is also used to correct small discrepancies in the transverse momenta spectrum of the ZZ system, using an event-by-event weight computed from data over simulation in the sideband~\cite{ref:2l2q}. 

In ggH-like events, the $\mathrm{t\bar{t}}$ background is estimated from data using e$^{\pm}\mu^{\mp}$ events passing the selection described above, which are dominated by top quark pair production events. The validity of this data-driven method is confirmed using a top-enriched sample, requiring events outside the dilepton mass window and with high $\cancel{E}_{T}$ ($\lambda>8$). In VBF-like events, the $\mathrm{t\bar{t}}$ production plays a minor role and it is estimated using a simulated sample generated using PYTHIA.

The diboson production is a small contribution to the total background. It is estimated using simulated samples, generated using Pythia. 

The signal is extracted from POWHEG simulated events. The signal simulation is modified to take into account the correct H lineshape, using the Complex-Pole Scheme approximation~\cite{ref:goria}, important for Higgs bosons with masses above 400~GeV. Moreover, the effect of the interference with the gg$\rightarrow$ZZ~\cite{ref:passarino} process is also included in the ggH simulation.

\section{Results}

Several systematic uncertainties, affecting both the shape and normalization of the signal and background, are included in the result. The most important systematic effects from lepton reconstruction are the muon momentum scale, electron energy scale and efficiency on lepton identification, isolation and triggering. From jet reconstruction the main uncertainties are: the jet energy scale, pile up, jet flavor tagging and boosted Z tagging. The following theoretical systematics are also considered: H production mechanism, signal cross section and branching fraction uncertainties. The background determination methods have the following uncertainties associated: Z+jets $p_{T}$(ZZ) correction uncertainty, the limited statistics in the e$^{\pm}\mu^{\mp}$ events and the diboson cross section uncertainty.

\begin{figure}
\begin{minipage}{0.33\linewidth}
\centerline{\includegraphics[width=1\linewidth]{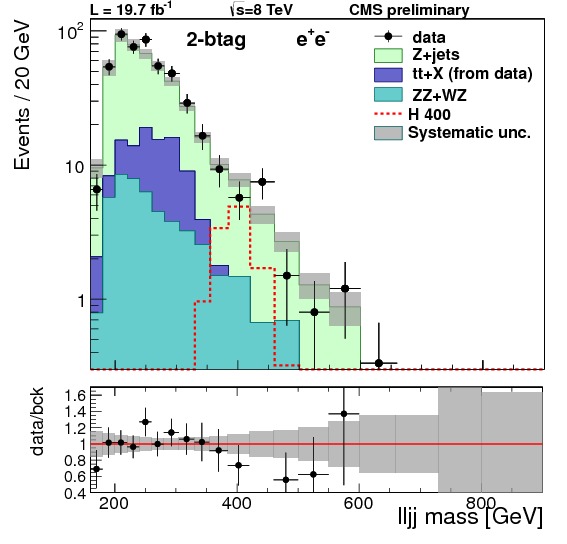}}
\end{minipage}
\hfill
\begin{minipage}{0.32\linewidth}
\centerline{\includegraphics[width=1\linewidth]{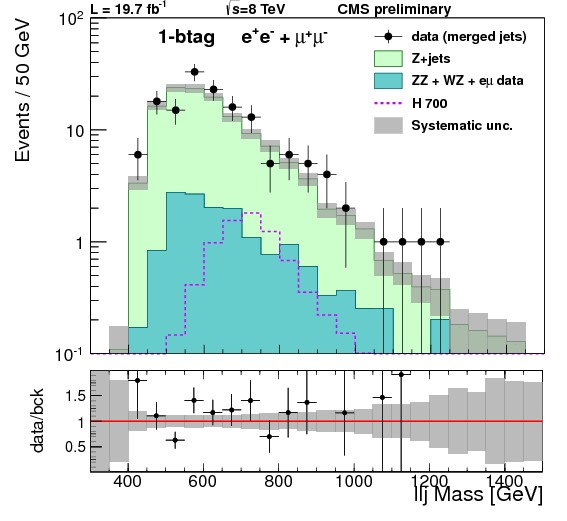}}
\end{minipage}
\hfill
\begin{minipage}{0.32\linewidth}
\centerline{\includegraphics[width=1\linewidth]{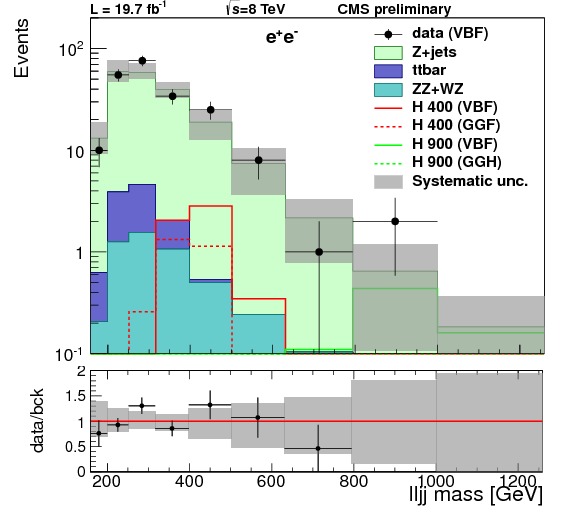}}
\end{minipage}
\caption[]{$m_{\ell\ell q\bar{q}}$ distribution for three categories of the analysis: dielectron dijet 2-btag (left), merged-jet 1-btag (center) and dielectron VBF (right).}
\label{fig:mZZ}
\end{figure}

\begin{figure}
\centering
\begin{minipage}{0.33\linewidth}
\centerline{\includegraphics[width=1\linewidth]{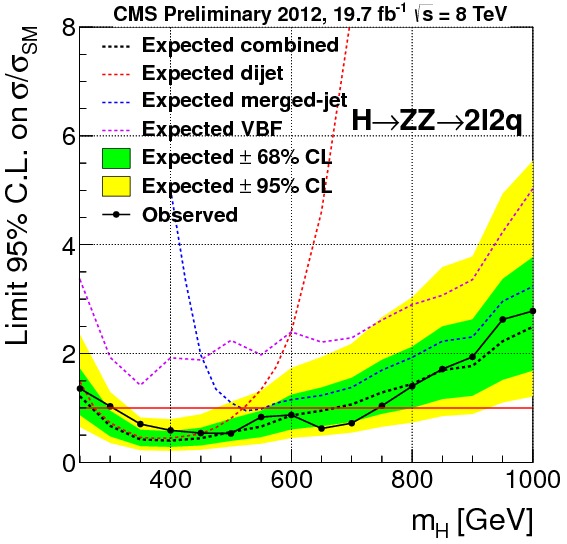}}
\end{minipage}
%\hfill
\begin{minipage}{0.33\linewidth}
\centerline{\includegraphics[width=1\linewidth]{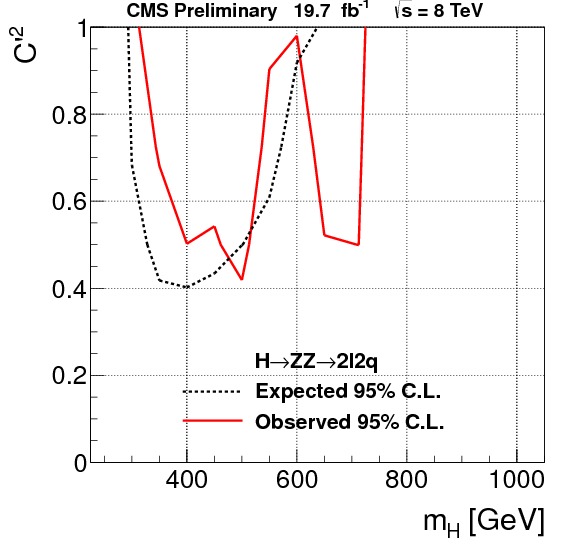}}
\end{minipage}
\caption[]{(Left) 95\% C.L. upper limits on the signal strength for the SM-like Higgs boson as a function of its mass. (Right) Observed and expected 95\% C.L. exclusion limits on the $C'^2$ parameter of the electroweak singlet model as a function of the H mass, under the $B_{new}=0$ hypothesis.}
\label{fig:limits}
\end{figure}

The $m_{\ell\ell q\bar{q}}$ distribution is studied for the 14 categories separately. As an example, three of these spectra are shown in Fig.~\ref{fig:mZZ}, which show an overall good agreement between data and expectation. The data are compared to the signal-plus-background and background-only distributions for 16 heavy SM-like Higgs boson mass hypothesis, ranging from 230 to 1000~GeV. There is no significant deviation of data with respect to the background expectations. Based on $m_{\ell\ell q\bar{q}}$ distributions, a statistical maximum likelihood fit is performed simultaneously in the 14 categories of the analysis, using the modified frequentist CL$_s$ approach~\cite{ref:stat}, treating systematic uncertainties as nuisance parameters. As no significant deviation of data with respect to the background is observed, exclusion limits on the signal strength, $\mu=\sigma/\sigma_{SM}$, are calculated (Fig.~\ref{fig:limits}, left). This analysis excludes the existence of a SM-like Higgs boson in the mass range between 305 and 744~GeV. 

Additionally, the results are interpreted in a model with an additional electroweak singlet mixed with the SM Higgs boson. The couplings of the SM Higgs boson and the electroweak singlet are modified by the scale factors $C$ and $C'$~\cite{ref:2l2q}. The unitarity is preserved by imposing the condition $C^2+C'^2=1$. The signal strength $\mu'$ and width $\Gamma'$ of the electroweak singlet are modified with respect to the SM predictions by:
\bea
\mu'     & = C'^2\cdot(1-B_{new}) \nonumber\\
\Gamma'  & = \Gamma_{SM}\cdot\frac{C'^2}{1-B_{new}}~,\nonumber
\label{eq:ewk}
\eea
where $B_{new}$ is the branching fraction of the electroweak singlet to decay modes not predicted by the SM. The analysis is performed similarly to the SM-like Higgs boson interpretation, but weighting the simulated samples to match the width and signal strength defined above. The expected and observed excluded region in the $C'^2$-$m_H$ plane is shown in Fig.~\ref{fig:limits} (right), for the $B_{new}=0$ case.

The results presented in this work are included in the CMS combination of H$\rightarrow$ZZ and H$\rightarrow$WW channels~\cite{ref:comb}, which excludes the presence of a SM-like Higgs boson up to $m_H=$1~TeV and provides a strong constraint in the $C'^2$-$B_{new}$ parameter space of the electroweak singlet interpretation.

\end{document}